\documentclass{rhumj_new}

\title[Degeneracies In a Wighted Sum of Two Squares]{Degeneracies In a Weighted Sum of Two Squares}

\author[Ishan Vinayagam Ramesh]{Ishan Vinayagam Ramesh}
\affiliation{Hopkington High School, Hopkington, Massachusetts 01748, USA}
\author{Maxim Olshanii\thanks{
We are grateful to Nathan Harshman, Danshyl Boodhoo, and Vanja Dunjko for valuable comments.}}
\affiliation{Department of Physics, University of Massachusetts Boston, Boston Massachusetts 02125, USA}
\email{Maxim.Olchanyi@umb.edu}


\begin{document}
\abstract{
This work is an attempt to classify and quantify instances when a weighted sum of two squares of positive integers, $3n_{1}^2+n_{2}^2$, can be realized in more than one way. Our project was inspired by a particular study of two-dimensional quantum billiards 
[S. G. Jackson, H. Perrin, G. E. Astrakharchik, and M. Olshanii, \emph{SciPost Phys. Core} {\bf 7}, 062 (2024)] where the weighted sums of interest represents an energy level with the two integers being the billiard's quantum numbers; there, the 3-fold degeneracies 
seem to dominate the energy spectrum. Interestingly, contrary to the conventional paradigm, these degeneracies are not caused 
by some non-commuting symmetries of the system.}
% Ishan V. Ramesh Edits: I made some minor grammatical corrections to the abstract above.

%
%==================================
\section{Introduction}
%
%
%%%%%%%%%%%%%%%%%%
\subsection{Background and statement of purpose}
% Ishan V. Ramesh Edits: I changed the subsection title from "Motivation for this project and its goal" to this and made some minor grammatical edits to the paragraph below.
%
This project was inspired by a particular study of two-dimensional quantum billiards perturbed by a diagonal $\delta$-function wall, which in turn, was a part of a research program \cite{jackson2024_062} devoted to generalizations of Bethe Ansatz \cite{gaudin1983_book_english}---a powerful method for solving a class of many-body quantum problems. In \cite{jackson2024_062} , it has been observed that many of the quantum energy levels of a $1\!:\sqrt{3}$ aspect ratio rectangular billiard without a $\delta$-ridge are triply \emph{degenerate}, i.e. they are supported by three quantum states simultaneously. While these triple degeneracies were successfully mathematically explained in \cite{jackson2024_062}, it remained an open question of if there are other systematic degeneracies and what their mathematical nature is. 

In the context of the underlying physical systems, quantum degeneracies are traditionally regarded as signs of the presence of \emph{non-commuting} symmetries in a problem~\footnote{In the context of billiards, consider all the transformations of space (rotations and mirror reflections) that map the billiard to % Ishan V. Ramesh Edits: In the text component right before and after, I made some minor grammatical edits (CONTINUE EDITING FROM HERE).
itself. Two of these transformations will be non-commuting if the outcome of their successive application depends on the order in which 
they were applied.}, with the Coulomb problem and the three-dimensional harmonic oscillator quoted as prime examples \cite{landau_quantum}(\S36). 
A notable exception is the so-called ``accidental degeneracies'' in a two-dimensional square box; many can be shown to be associated with properties of the Hamiltonian under dilations, as opposed to reflections, the former 
being further related to the number-theoretical properties of the spectrum \cite{shaw1974_1537}.
These respective appearances of numerous systematic degeneracies in a rectangular billiard---whose symmetry operations do commute---need to first be studied and quantified before eventually, are understood.  % Ishan V. Ramesh Edits: I made some grammatical edits in the paragraph prior. Original final sentence of the paragraph is shown below.
% "In this respect appearance of the numerous systematic degeneracies in a rectangular billiard---whose symmetry operations do commute---needs to be first studied and quantified and then, eventually, understood."

%
%%%%%%%%%%%%%%%%%%
\subsection{Glossary of terms}
While the question of multiplicity of a representation of an integer as a weighted sum of two squares of integers naturally falls under fields of number theory, it is necessary to preserve the language of quantum systems the question originates from, both because it appears to be suitably succinct and of relevance to future studies in which one may decide to eventually return to the quantum billiards using our results. We suggest the following 
nomenclature: % Ishan V. Ramesh Edits: I edited the previous paragraph for some grammatical misprints.
\begin{itemize}
\item[] \emph{State}: an ordered pair of positive integers, $(n_{1},\, n_{2})$, e.g.\ $(1,\, 5)$. This is the analogue of a quantum eigenstate of the 
original problem \cite{jackson2024_062}; 
\item[] \emph{State's indices}: the integers $n_{1}$ and $n_{2}$(, e.g.\  $1$ and $5$, in case of the example just above). They correspond to the quantum numbers of the original 
problem \cite{jackson2024_062}; 
\item[] \emph{Energy of a state}: the weighted sum of squares of interest evaluated on the state's indices, 
$\mathcal{E}_{n_{1},n_{2}} \equiv 3n_{1}^2+n_{2}^2$. It directly corresponds to the energy spectrum of \cite{jackson2024_062} if the latter is measured in the units of $\frac{\pi^2}{6} \frac{\hbar^2}{m L^2}$, where $L$ is the width of the billiard ($\sqrt{3} L$ being its height), $m$ is the mass of the particle in it, and $\hbar$  is the Planck's constant. The energy for the the example above is 
$\mathcal{E}_{1,\,5} = 3\cdot1^2+5^2=28$;    
\item[] \emph{Energy level}: a set of all states of a given energy $\mathcal{E}$: \\
$\Big\{(n_{1}^{(1)},\, n_{2}^{(1)}),\,(n_{1}^{(2)},\, n_{2}^{(2)}),\,\ldots,\,(n_{1}^{(g)},\, n_{2}^{(g)})\Big\}$, where
$3(n_{1}^{(1)})^2+(n_{2}^{(1)})^2=3(n_{1}^{(2)})^2+(n_{2}^{(2)})^2= \ldots = 3(n_{1}^{(g)})^2+(n_{2}^{(g)})^2=\mathcal{E}$. The energies $\mathcal{E}$ of the energy levels can be used to label them. In this text, we will be using large braces, 
$\Big\{\cdots\Big\}$, to denote the whole set of states comprising an energy level. Example: $\Big\{(1,\, 5),\,(2,\, 4),\,(3,\, 1)\Big\}$ comprise the energy level $\mathcal{E}=28$. Indeed, 
$3\cdot1^2+5^2=3\cdot2^2+4^2=3\cdot3^2+1^2=28$;
\item[] \emph{Degree of degeneracy of an energy level}: the number of states the energy level consists of; the number $g$ in the example above. An energy level $\mathcal{E}$ is called \emph{degenerate} if its degree of degeneracy is greater than one. 

\item[] \emph{Energy spectrum}: the full set of energies labeled by the their corresponding states: 
$\mathcal{E}_{n_{1},n_{2}}$.
The energy spectrum is called \emph{degenerate} if it features at least one 
degenerate energy level; 
\item[] \emph{Multiplets}: this notion is specific to our project, and it is \emph{not} related to the parental problem \cite{jackson2024_062}. A \emph{mutiplet} is a subset of the states of the same energy (and as such, belonging to the 
same energy level)  predicted mathematically. We will be focusing on the \emph{triplets} (three-member multiplets) and  \emph{doublets} (two-member multiplets). Example: energy level 
$\mathcal{E}=196$ contains four states, $\Big\{(3,\, 13),\,(5,\, 11),\,(8,\, 2),\,(7,\, 7)\Big\}$. The first three, 
$\{(3,\, 13),\,(5,\, 11),\,(8,\, 2)\}$ are a triplet predicted by the Perrin's formula \cite{jackson2024_062} described in detail far below: $\{(m_{1},\,m_{1}+2m_{2}),\,(m_{2},\,m_{2}+2m_{1}),\,(m_{2}+m_{1},\,m_{2}-m_{1})\}$ with $m_{1}=3$ and $m_{2}=5$. In this text, we will be using regular braces, 
$\{\cdots\}$, to denote mathematically predicted multiplets of states of the same energy.
\end{itemize}
%

%%%%%%%%%%%%%%%%%%
\subsection{Statement of the problem}
The object of this study is  the energy spectrum of an  aspect ratio $1\!:\!\sqrt{3}$ rectangular quantum billiard \cite{jackson2024_062}:
\begin{align}
\begin{split}
&
\mathcal{E}_{n_{1},n_{2} } = 3n_{1}^2+n_{2}^2 
\\
&
\text{with}
\\
&
n_{1}=1,\,2,\,\ldots
\\
&
n_{2}=1,\,2,\,\ldots
\end{split}
\label{spectrum}
\end{align}
It has been numerically observed \cite{jackson2024_062} that this spectrum features degenerate energy levels, with 3-fold degeneracies occupying a substantial portion of the spectrum. The first such degeneracy occurs at $\mathcal{E}=28$; the states that comprise this energy level are $\Big\{(1,\, 5),\,(2,\, 4),\,(3,\, 1)\Big\}$. Our goal is to describe this phenomenon mathematically, find other prominent degeneracies, and prepare grounds for identifying the physical mechanisms behind the phenomenon. % Ishan V. Ramesh Edits: Changed "understand" to "describe".   

%
%%%%%%%%%%%%%%%%%%
\subsection{State of the art}
\subsubsection{Perrin triplets}
In \cite{jackson2024_062}, it has been found that the 3-fold degeneracies in the spectrum 
\eqref{spectrum} systematically occur at energies $\mathcal{E}=4(m_{1}^2+m_{1}m_{2}+m_{2}^2)$. Using this observation, 
Perrin and others \cite{jackson2024_062} were able to produce a formula that generates all three states of that energy level, for any pair  of 
$m_{1},\,m_{2}$:
% Ishan V. Ramesh Edits: I changed "energy" to "energy level" in the preceding plain text portion concerning 1.4.1 Perrin triplets to keep a concise vocabulary consistent with the glossary of terms.
%
\begin{align}
\begin{split}
&
\mathcal{E}_{n_{1}^{(\text{P})},n_{2}^{(\text{P})} } = 3( n_{1}^{(\text{P})} )^2+(n_{2}^{(\text{P})})^2 =4(m_{1}^2+m_{1}m_{2}+m_{2}^2)
\\
&
(n_{1}^{(\text{P})},\,n_{2}^{(\text{P})}) \in \{(n_{1}^{(\text{P,1})},\,n_{2}^{(\text{P,1})}),\,(n_{1}^{(\text{P,2})},\,n_{2}^{(\text{P,2})}),\,(n_{1}^{(\text{P,3})},\,n_{2}^{(\text{P,3})})\}
\\
&
\text{with}
\\
&
(n_{1}^{(\text{P,1})},\,n_{2}^{(\text{P,1})})=(m_{1},\,m_{1}+2m_{2})
\\
&
(n_{1}^{(\text{P,2})},\,n_{2}^{(\text{P,2})})=(m_{2},\,m_{2}+2m_{1})
\\
&
(n_{1}^{(\text{P,3})},\,n_{2}^{(\text{P,3})})=(m_{2}+m_{1},\,m_{2}-m_{1})
\\
&
\text{where}
\\
&
m_{1}=1,\,2,\,3,\,\ldots
\\
&
m_{2}=m_{1}+1,\,m_{1}+2,\,m_{1}+3,\,\ldots
\end{split}
\label{Perrin}
\end{align}
\subsubsection{Brahmagupta doublets}
In the first half of the 7th century BC, Brahmagupta produced what is now recognized as Brahmagupta identity
% Ishan V. Ramesh Edits: I changed the syntax in the second half of the sentence from "what will be later called" to "what is now recognized as".
\cite{suryanarayan1996_30}:
\begin{align}
\begin{split}
(m \nu_{1}^2+\nu_{2}^2)(m \nu_{3}^2+\nu_{4}^2)
=&
m (\nu_{1}\nu_{4}-\nu_{2}\nu_{3})^2 + (m\nu_{1}\nu_{3}+\nu_{2}\nu_{4})^2
\\
=&
m (\nu_{1}\nu_{4}+\nu_{2}\nu_{3})^2 + (m\nu_{1}\nu_{3}-\nu_{2}\nu_{4})^2
\end{split}
\,\,,
\label{B_formula}
\end{align}
where $m$ and $\nu_{\cdot}$ are arbitrary real numbers. Brahmagupta's formula was intended to prove that a product of two weighted sums of two % Ishan V. Ramesh Edits: Made some minor grammatical adjustments to the sentence above.
squares of integers  $m l_{1}^2 + l_{2}^2$ is also a weighted sum sum of two integers with the same weight $m$. The $m=1$ instance
of this identity (known as Diophantus-Brahmagupta-Fibonacci identity) 
appears in  Diophantus' {\it Arithmetica} \cite{diophantus_arithmetica} (3-rd century AD).

In the context of the present article, it is important to notice that at $m=3$ and for integer values of the 
$\nu_{1}\nu_{4}\pm\nu_{2}\nu_{3}$ and $3\nu_{1}\nu_{3}\pm\nu_{2}\nu_{4}$ combinations, the Brahmagupta identity 
\eqref{B_formula} produces doublets of the energy-degenerate states of our problem \eqref{spectrum}:
\begin{align}
\begin{split}
&
\mathcal{E}_{n_{1}^{(\text{B})},n_{2}^{(\text{B})} }  = 3( n_{1}^{(\text{B})} )^2+(n_{2}^{(\text{B})})^2 = (3\nu_{1}^2+\nu_{2}^2)(3\nu_{3}^2+\nu_{4}^2)
\\
&
(n_{1}^{(\text{B})},\,n_{2}^{(\text{B})}) \in \{(n_{1}^{(\text{B,1})},\,n_{2}^{(\text{B,1})}),\,(n_{1}^{(\text{B,2})},\,n_{2}^{(\text{B,2})})\}
\\
&
\text{with}
\\
&
(n_{1}^{(\text{B,1})},\,n_{2}^{(\text{B,1})})=(|\nu_{1}\nu_{4}-\nu_{2}\nu_{3}|,\,3\nu_{1}\nu_{3}+\nu_{2}\nu_{4})
\\
&
(n_{1}^{(\text{B,2})},\,n_{2}^{(\text{B,2})})=(\nu_{1}\nu_{4}+\nu_{2}\nu_{3},\,|3\nu_{1}\nu_{3}-\nu_{2}\nu_{4}|)
\\
&
\text{where $\nu_{1}$, $\nu_{2}$, $\nu_{3}$, $\nu_{4}$ are such that}
\\
&
n_{\ldots}^{(\text{B}, \cdots)}(\nu_{1}, \nu_{2}, \nu_{3}, \nu_{4}) \text{ are positive integers}
\end{split}
\label{Brahmagupta}
\end{align}
%

%
%==================================
\section{Results}
\subsection{Parity of the states comprising a degenerate manifold}
In this subsection, we are going to look at the relative parity of the indices $n_{1}$ and $n_{2}$ for the states $(n_{1},\,n_{2})$ comprising a degenerate manifold of an energy level \[\mathcal{E}=\mathcal{E}_{n_{1},n_{2}}=3n_{1}^2+n_{2}^2\,\,.\]
% Ishan V. Ramesh Edits: To ensure a concise vocabulary, I changed "energy" to "energy level" in the plain text section under 2.1 above.
Observe that for any even value of $\mathcal{E}$, an odd $n_{1}$ would imply an odd $n_{2}$, and an even $n_{1}$ would imply an even $n_{2}$. Notice that this property would be held by all the states in the manifold. At the same time, for an odd $\mathcal{E}$, $n_{1}$ and $n_{2}$ in all constituent states must have an opposite parity. This observation allows us to classify the degenerate manifolds by the relative parity of the indices  $n_{1}$ and $n_{2}$ in the states the manifold consists of. We can introduce the following \emph{definition}:
\begin{itemize}
\item[]
A degenerate manifold is called a \emph{same(opposite)-parity degenerate manifold} if for all the states in the manifold, $n_{1}$ and $n_{2}$ have the same(opposite) parity. 
\end{itemize}
And we have just shown that 
\begin{itemize}
\item[]
any degenerate manifold is either a same-parity manifold or an opposite-parity manifold. 
\end{itemize}

The above definition can immediately prove useful. For example, one can easily show that 
a Perrin triplet can be only contained in a same-parity degenerate manifold.  % Ishan V. Ramesh: Earlier, "degenerate" was printed twice.

\subsection{Every pair of energy-degenerate states is a Brahmagupta doublet \label{ssec:everything_is_a_brahmagupta_doublet}}

Here, we are going to show that every pair of energy-degenerate states is a Brahmagupta doublet. Indeed,  any two states $(\tilde{n}^{(1)}_{1},\,\tilde{n}^{(1)}_{2})$ and $(\tilde{n}^{(2)}_{1},\,\tilde{n}^{(2)}_{2})$ 
of the same energy, i.e.\ such that 
\begin{align}
3(\tilde{n}^{(1)}_{1})^2+(\tilde{n}^{(1)}_{2})^2 = 3(\tilde{n}^{(2)}_{1})^2+(\tilde{n}^{(2)}_{2})^2
\label{intermediate_1}
\end{align}
form
a Brahmagupta doublet with 
\begin{align}
\begin{split}
&
\tilde{\nu}_{1}=(\tilde{n}^{(2)}_{1}+\tilde{n}^{(1)}_{2})\xi
\\
&
\tilde{\nu}_{2}=3(\tilde{n}^{(2)}_{1}-\tilde{n}^{(1)}_{1})\xi
\\
&
\tilde{\nu}_{3}=\frac{1}{6\xi}
\\
&
\tilde{\nu}_{4}=\frac{\tilde{n}^{(2)}_{1}+\tilde{n}^{(1)}_{1}}{2(\tilde{n}^{(2)}_{2}+\tilde{n}^{(1)}_{2})\xi}
\end{split}
\,\,.
\label{B_inverse_1}
\end{align}
and
\begin{align}
\begin{split}
&
(\tilde{n}_{1}^{(\text{1})},\,\tilde{n}_{2}^{(\text{1})})=(\tilde{\nu}_{1}\tilde{\nu}_{4}-\tilde{\nu}_{2}\tilde{\nu}_{3},\,3\tilde{\nu}_{1}\tilde{\nu}_{3}+\tilde{\nu}_{2}\tilde{\nu}_{4})
\\
&
(\tilde{n}_{1}^{(\text{2})},\,\tilde{n}_{2}^{(\text{2})})=(\tilde{\nu}_{1}\tilde{\nu}_{4}+\tilde{\nu}_{2}\tilde{\nu}_{3},\,3\tilde{\nu}_{1}\tilde{\nu}_{3}-\tilde{\nu}_{2}\tilde{\nu}_{4})
\end{split}
\quad.
\label{B_inverse_2}
\end{align}
Notice that this connection is identical to the association \eqref{Brahmagupta} with the absolute value sign being resolved to a plus sign. Recall that in this construction, we assume all four indices $(\tilde{n}^{(1)}_{1},\,\tilde{n}^{(1)}_{2})$ and $(\tilde{n}^{(2)}_{1},\,\tilde{n}^{(2)}_{2})$ are positive. % Ishan V. Ramesh Edits: Made a minor grammatical edit in the previous sentence.

\subsection{Numerical results \label{ssec:numerical}}
In this subsection, we have performed a small numerical investigation of the  spectrum \eqref{spectrum}. We included all energy levels with energies % Ishan V. Ramesh Edits: I inserted "In this subsection" at the start of the first sentence of subsection 2.3.
less or equal to $\mathcal{E}_{\text{max}}=2700$. This energy range contains $1179$ states distributed over $655$ energy levels.
A detailed breakdown is presented in Table~\ref{table:1}.
\begin{table}[h!]
\centering
\begin{tabular}{||c c c ||} 
 \hline
 degree of degeneracy & \# of energy levels  & \# of states\\ [0.5ex] 
 \hline\hline
 \multicolumn{3}{||c||}{same-parity energy levels} \\
 1 & 32 & 32\\
 2 & 0 & 0 \\
 3 & 132 & 396 \\
 4 & 8 & 32 \\
 5 & 0 & 0 \\
 6 & 20 & 120 \\
 7 & 0 & 0 \\
 8 & 0 & 0 \\
 9 & 1 & 9 \\
 $>$9 & 0 & 0 \\
 \bf{sub-total} & \bf{193} & \bf{589}\\
 \hline 
 \multicolumn{3}{||c||}{opposite-parity energy levels} \\
 1 & 344 & 344 \\
 2 & 109 & 218 \\
 3 & 8 & 24 \\
 4 & 1 & 4 \\ 
 $>$4 & 0 & 0 \\ 
 {\bf sub-total} & {\bf 462} & \bf{590} \\ 
 \hline
 {\bf total} & {\bf 655} & \bf{1179} \\  
 \hline
\end{tabular}
\caption{Number of degenerate manifolds of each parity and degree of degeneracy and the corresponding number of states, for all energy levels \eqref{spectrum}
of energy below or equal  $\mathcal{E}_{\text{max}}=2700$.}
\label{table:1}
\end{table}
From this Table, one can immediately see that % Ishan V. Ramesh Edits: I changed "that" to "this" in the sentence's (in this line) second word and made some grammatical edits in lines 363, 364, & 366.
\begin{itemize}
\item[(a)] the same-parity energy levels are dominated by the 3-fold degenerate energy levels. The first instance 
of such an energy level is $\mathcal{E}=28$, which is populated by $\Big\{(1,\, 5),\,(2,\, 4),\,(3,\,1)\Big\}$;
\item[(b)] the opposite-parity  2-fold degeneracies dominate the set of the opposite-parity degenerate energy levels. The first instance 
of such an energy level is $\mathcal{E}=91$, which is populated by $\Big\{(3,\, 8),\,(5,\, 4)\Big\}$.
\end{itemize}
Further investigation shows that within the energy range studied, $\mathcal{E}\le 2700$, resultingly, % Is adding "resultingly" okay here.
\begin{itemize}
\item[(a)] all same-parity 3-fold degenerate energy levels are of the Perrin type \eqref{Perrin}. For example, the states 
$\Big\{(1,\, 5),\,(2,\, 4),\,(3,\,1)\Big\}$ comprising the $E=28$ level can be interpreted as a Perrin triplet  \eqref{Perrin} with 
$(m_{1}, \,m_{2})=(1,\,2)$;
\item[(b)] all opposite-parity 2-fold degenerate energy levels are given by the Brahmagupta formula 
\eqref{Brahmagupta} with $\nu_{1}$ and $\nu_{2}$ being positive integers and $\nu_{3}$ and $\nu_{4}$ 
being positive half-integers. The energy levels below $1267$ allow both half-integer and integer representations. E.g.\ 
the lowest opposite-parity 2-fold degenerate level $\mathcal{E}=91$ can be Brahmagupta-represented (see \eqref{Brahmagupta})
in 16 distinct ways, two of which are all integer: 
$(\nu_{1},\,\nu_{2},\,\nu_{3},\,\nu_{4})=
(1,\, 2,\, 2,\, 1),\, (2,\, 1,\, 1,\, 2),\, (1,\, 1,\, \frac{3}{2},\, 4),\, (1,\, 1,\, \frac{5}{2},\, 2),\, (1,\, 2,\, \frac{1}{2},\, \frac{7}{2}),\, \\ (1,\, 2,\, \frac{3}{2},\, \frac{5}{2}),\,  (1,\, 5,\, 1,\, \frac{1}{2},\, 91),\, (1,\, 7,\, \frac{1}{2},\, 1),\, (2,\, 1,\, \frac{1}{2},\, \frac{5}{2}),\,  (2,\, 1,\, \frac{3}{2},\, \frac{1}{2}),\,  (2,\, 4,\, 1,\, \frac{1}{2}),\, (3,\, 1,\, 1,\, \frac{1}{2}),\, \\(3,\, 5,\, \frac{1}{2},\, 1),\, (3,\, 8,\, \frac{1}{2},\, \frac{1}{2}),\, (4,\, 2,\, \frac{1}{2},\, 1),\, (5,\, 4,\, \frac{1}{2},\, \frac{1}{2})$. The lowest opposite-parity 2-fold degenerate energy level that does 
require half-integers is $\mathcal{E}=1267$; its six Brahmagupta representations \eqref{Brahmagupta}) are:
$
(\nu_{1},\,\nu_{2},\,\nu_{3},\,\nu_{4})=
(1,\,  1,\,  \frac{17}{2},\,  10),\,  (1,\,  2,\,  11/2,\,  \frac{19}{2}),\,  \\(1,\,  2,\,  \frac{15}{2},\,  7/2),\,  (1,\,  5,\,  1,\,  \frac{13}{2}),\,  (2,\,  4,\,  1,\,  \frac{13}{2}),\,  (3,\,  1,\,  1,\,  \frac{13}{2})
$.

Note that while according to the Subsection~\ref{ssec:everything_is_a_brahmagupta_doublet}, any pair of energy-degenerate states forms a Brahmagupta doublet, the integer/half-integer nature of the $\nu$-indices is not a priori guaranteed. % Ishan V. Ramesh Edits: Is the change from "priori guaranteed" a correct change?
\end{itemize}
%

%==================================
\section{Summary of results}
% Is "Rectangular Quantum Billiard" right after the aspect ratio number grammatically correct.
In this article, we have addressed the energy degeneracies in the energy spectrum of an aspect ratio $1\!:\!\sqrt{3}$ rectangular quantum billiard. We found that all states $(n_{1},\,n_{2})$ comprising a given energy level possess the same relative parity of the state indices (i.e.\ quantum numbers) $n_{1}$ and $n_{2}$. Consequently, the relative parity of the constituent states---that happens to coincide 
with the parity of the energy of the level---constitutes an ideal candidate for the top level taxonomy of the energy levels. 
% Ishan V. Ramesh Edits: In the above paragraph, I changed "induces" to "indices" (I believe that "indices" was the intended word).

Despite the simplicity of the symmetry properties of this billiard, we found, numerically, two systematic series of the degenerate energy levels. 

% Ishan V. Ramesh Edits: Made some minor grammatical edits below and changed the paragraph slightly, the original paragraph is shown below:
% "Remark also that numerically, the Perrin series
% appears to dominate the same-parity portion of the energy spectrum: 132 out of the
% available 193 energy levels were found to be of the Perrin nature."

For the levels of the same-parity type, all the 3-fold degenerate levels are represented (within the 
numerically studied range of energies) by the Perrin series \eqref{Perrin}. We conjectured that this was a general phenomenon. It is also shown that, numerically, the Perrin series appears to dominate the same-parity portion of the energy spectrum as 132 out of the 
available 193 energy levels were found to be of the Perrin nature.  

% Ishan V. Ramesh Edits: I made some minor edits to the sentence below.
For levels of the opposite-parity type, there exists a series of 2-fold degenerate levels represented by the Brahmagupta doublets \eqref{Brahmagupta} with integer and semi-integer entries. This series is less dominant than the Perrin one: it features 109 levels out of 462 available. However, it is still prominent among the \emph{degenerate} energy levels, with only 9 degenerate levels not being manifestly of the Brahmagupta type.    

%==================================
\section{Directions of future study}
%

% Ishan V. Ramesh Edits: In most of the section text blocks of the "Future projects" section as well as the titling of the section itself, I made some minor edits.
Our numerical results outlined in the Subsection~\ref{ssec:numerical} inspire two directions of further research delegated to the future study.
\begin{itemize}
\item[(a)] In the future, we would like to attempt to prove that \emph{all} same-parity 3-fold degenerate energy levels show 
the Perrin structure \eqref{Perrin}; 
\item[(b)] A much more difficult future project is (i) to prove that \emph{all} opposite-parity 2-fold degenerate energy levels 
possess a Brahmagupta representation \eqref{Brahmagupta} with $\nu_{1}$ and $\nu_{2}$ being positive integers and $\nu_{3}$ and $\nu_{4}$ being positive semi-integers and, subsequently (ii) identify an indexing system that establishes a one-to-one correspondence 
between the opposite-parity 2-fold degeneracies and the four-member strings of the $\nu$ indices, $(\nu_{1},\,\nu_{2},\,\nu_{3},\,\nu_{4})$ (similar to the map \eqref{Perrin} for the same-parity 3-fold degeneracies and the pairs $(m_{1}, \,m_{2})$).
\end{itemize}

Our long-term plan is to understand the \emph{physical} origins of the degeneracies in the spectrum \eqref{spectrum}. In doing so, we will attempt to step beyond the current paradigm of degeneracies being caused by 
non-commuting underlying symmetries \cite{landau_quantum,shaw1974_1537} and to try to discover new physical mechanisms that cause degeneracies. 

At the moment, odd-parity 2-fold degeneracies constitute a complete mystery from a physics perspective, and we plan to have this question as our first objective of further study.

However, the 3-fold same-parity degeneracies offer a hint. Their energies,
\begin{align*}
\begin{split}
&
\mathcal{E}=4(m_{1}^2+m_{1}m_{2}+m_{2}^2)
\\
&
m_{1}=1,\,2,\,3,\,\ldots
\\
&
m_{2}=m_{1}+1,\,m_{1}+2,\,m_{1}+3,\,\ldots
\end{split}
\,\,,
\end{align*}
happen to coincide with the energy spectrum the 
$30^{\circ}-60^{\circ}-90^{\circ}$ triangle that tiles the original rectangle via  reflections about its sides 
(see Fig.\ 8 in \cite{jackson2024_062}). While this triangular billiard can indeed be treated using 
the symmetries of a hexagon that do support degeneracies (see \cite{olshanii2015_105005} for a general approach to solvable 
simplex-shaped billiards), the spectrum of the  $30^{\circ}-60^{\circ}-90^{\circ}$ triangular billiard per se is not degenerate. Yet, we believe 
% Ishan V. Ramesh Edits: I made some minor grammatical edits to the sentence below.
that this connection will be proven to be a key to an understanding of the physical reasons for the existence of multiple 3-fold divergencies 
in our system. In this context, the general question of whether systematic degeneracies in the spectrum of a quantum billiard can be caused by the symmetries of the individual tiles in the parquet that covers that billiard appears to be worthy of future investigation.   
% Ishan V. Ramesh Edits: Above, I added "future" before "investigation as an edit.

\end{document}